**Classification: Physical Sciences (Applied Physical Sciences), Biological Sciences (Ecology)**

**Title: Similar self-organizing scale-invariant properties characterize early cancer invasion and long range species spread**

Diana E. Marco[*], Sergio A. Cannas[†], Marcelo A. Montemurro[‡], Bo Hu[§, ¶] and Shi-Yuan Cheng[§, ¶¶]

[*]Laboratorio de Ecología Matemática, Area de Producción Orgánica, Facultad de Ciencias Agropecuarias, Universidad Nacional de Córdoba, Ciudad Universitaria, CC 509, 5000 Córdoba, Argentina. [†]Facultad de Matemática, Astronomía y Física, Universidad Nacional de Córdoba, Ciudad Universitaria, 5000 Córdoba, Argentina. [‡]Faculty of Life Sciences, University of Manchester, Jackson's Mill, G7, PO Box 88, Sackville St, Manchester, M60 1QD, United Kingdom. [§]University of Pittsburgh Cancer Institute & Department of Medicine[¶] and Pathology[¶¶], Research Pavilion at the Hillman Cancer Center, Suite 2.26, 5117 Centre Avenue, Pittsburgh, PA 15213-1863, USA.

**Corresponding author:** Diana E. Marco, Laboratorio de Ecología, Area de Producción Orgánica, Facultad de Ciencias Agropecuarias,Universidad Nacional de Córdoba, Av. Valparaíso s/n, Ciudad Universitaria, CP 5000, CC 509, Tel. +54-351-4334103/05/16/17/18 ext. 255 Fax +54-351-4334103/05/16/17/18 ext. 114

e-mail: dmarco@agro.uncor.edu

**Manuscript information**

Abstract: 129 words

Text: estimated 32,016 characters including spaces.

**Abstract**

Occupancy of new habitats through dispersion is a central process in nature. In particular, long range dispersal is involved in the spread of species and epidemics, although it has not been previously related with cancer invasion, a process that involves spread to new tissues. We show that the early spread of cancer cells is similar to the species individuals spread and that both processes are represented by a common spatio-temporal signature, characterized by a particular fractal geometry of the boundaries of patches generated, and a power law-scaled, disrupted patch size distribution. We show that both properties are a direct result of long-distance dispersal, and that they reflect homologous ecological processes of population self-organization. Our results are significant for processes involving long-range dispersal like biological invasions, epidemics and cancer metastasis.

Long-distance dispersal (LDD) (1), even if represented by rare events, is one of the main factors explaining the fast spread of different organisms in new habitats, for example in paleocolonization events (2), plant pathogens (3), and invasive species (4). In addition, spread from primary tumors can be thought as a biological invasion from cancer cells spreading and invading new tissues. Considering cancer as an ecological process (5), the fitness of a neoplastic cell is shaped in part by its interactions with cells and other factors in its microenvironment, the surrounding tissues. Colonization begins with a single or few cells previously dispersed from the primary tumor (6), originating different clone lines that evolve independently across the new tissues and organs invaded (7) in a process we can consider as LDD. In spite of the remarkable similarity with species spread, at present no detailed mechanism has been proposed for an ecological interpretation of cancer spread.

In this paper we show that the spread of cells in cancer invasion and of invasive species generates a similar patchy pattern characterized by fractal and power law scaling. Furthermore, we show that this common pattern originates from self-organized, homologous mechanisms driven by LDD.

Fat-tailed functions like the power law seem to adequately describe LDD (1), and evidence for this is coming from crop pathogens distributions determined experimentally (8) and from model simulations (9,10). Power law functions and fractal geometry characterize species dispersal by LDD (9), and they reflect the invariance of some property over a range of temporal and spatial scales. There is increasing consensus in that they can be a byproduct of self-organizing processes of populations and communities (11, 12). The capacity of a system to evolve to an organized state due to intrinsic mechanisms, i.e., self-organization, often characterized by a scale free geometry, has been attributed to diverse natural phenomena (13). However, the fundamental dynamics that determine self-organization scaling properties have remained obscure in many cases. Performing independent simulations we show that the pattern properties we found in real data from cancer invasion and species spread are specific of long range dispersal mediated by a power law distribution function.

We performed simulations using a spatially explicit, individual-based model based on a cellular automaton originally developed for the study of biological invasions (14, 15). We simulated long ranged dispersal mechanisms using the power law dispersal function $f(r) = A / r^{\alpha}$ (9). The main biological significance of the inclusion of the power law in the model is that dispersion is allowed to reach the whole area considered without distance limits. This is drastically different respect to the use of distribution functions allowing only short-distance dispersal (SDD), where dispersion can reach just close areas to the initial focus (9). The power law function allows for the inclusion of local and LDD events in the same dispersal function, depending on the value of the α exponent (9). To characterize the spatial pattern of spread produced by the simulations, we calculated the mean fractal dimension $D$ of patch borders using a box counting algorithm (16), and determined the patch size distribution. We explored the pattern of spread produced by the model to understand the observed patterns of spread of cells from an invasive human glioma and of an invasive tree (English elm, *Ulmus minor* Mill.). We analyzed the spread of human glioma U87MG cells engineered to express an angiogenic regulator, angiopoietin-2 (Ang2) (U87MG/Ang2 cells), capable of promoting glioma cell infiltration into the brain parenchyma, established by intracranial cross-species transplantations in the brain of mice (17). Invasion of glioma cells involves the attachment of invading tumor cells to the extracellular matrix and its disruption, and subsequent invading cell penetration into brain tissues adjacent to primary tumor. This process is mediated by tumor-secreted enzymes called matrix metalloproteases (MMPs) that degrade the extracellular matrix at tumor-invasive borders and invasive areas (17). We recorded the spread of English elm into a native forest from an initial small focus using aerial photographs taken in 1970, 1987 and 1996. Fruits of English elm are dispersed by wind (usually assumed to be a LDD mechanism) in high numbers but many seeds remain near the parent providing also local dispersal.

**Results and discussion**

The analysis of pattern generation process with LDD during the simulations allows understanding its mechanism (Fig. 1a, Movie S1), which is essentially different from SDD mechanism (Movie S2, 9, 17 Fig. 2a). In LDD, beginning with an initially reproductive individual, a single patch appears surrounded by isolated, immature individuals (green dots, Movie S1) scattered all over the field. At times longer than the time of first reproduction, some of the scattered individuals begin reproduction (black dots, Movie S1), and secondary foci initiate growth into patches showing the same structure as the initial patch. While patches of higher order generation continue arising and growing, the first patch itself continues growing and absorbing the nearest patches. This coalescence of similar patches originates a self-similar border in the initial patch and it accounts for sudden increments in patch area. The same process of patch growth and coalescence occurs in other patches distant from the main patch. There is no a clearly defined invasion front (Movie S1; Fig. 1a), and the spatial extent of the population grows exponentially (Movie S1). In agreement with simulations results, spread of the invasive, Ang2-expressing gliomas displayed irregular borders with spike-like structures that invaded into the normal brain structures (Fig. 1b), in contrast with the non invasive control tumors showing smooth, clean borders (17 Fig. 2a) clearly resembling the SDD process (9, Movie S2). Glioma cells migrated far away the initial tumor and formed groups of individual tumor clusters that localized at 2.5–4.3 mm from the tumor mass in various invasive tumors, resembling the simulated patterns. Although it is not possible to image cell spreading in the brain tissue of mice in the *in vivo* model at sequential times, *in vitro* assays assessing the invasiveness of various glioma cells through membranes coated with Matrigel showed that U87MG/Ang2 cells had a 4-fold increase in invaded area compared with the parental U87MG cells (17), suggesting exponential increase.

The spatial pattern of spread of English elm from aerial photographs (Fig 1c) resembled closely the pattern obtained by simulations and glioma spread (Fig. 1 a, b). The temporal and spatial patterns of patch generation from the initial focus composed by a small patch present in 1970 explain the resulting similarities (not shown). The number of patches initially grew exponentially and then slowed down. In 1987, about 74 new patches covered a small fraction of the area and were mainly represented by individual trees scattered through the 7 hectares forest area. The increase in covered area was also exponential although faster and continued, with 189 patches present ten years later. The faster area growth rate reveals that after a certain time, few new patches are generated but the increase in area is mainly due to patch growth and coalescence, as occurred in the simulations (Movie S1).

The fractal dimension $D$ of patch borders from simulations as a function of $\alpha$ varied between 1.6 and 1.7 for $2 < \alpha \leq 3$, and it decreased monotonously for $3 < \alpha \leq 4$. We calculated the fractal dimension $D$ of patch borders from the digitized aerial images of English elm field cover in 1996. For the main field patch we found $D = 1.66$ ($R^2 = 0.93$, n = 10). Several of the remaining biggest patches showed $D$ values ranged between 1.40 and 1.75 ($R^2$ between 0.95 and 0.99, n = 10 in all cases). These values correspond to predicted power law $\alpha$ exponents for LDD distribution functions (9). In particular, $D = 1.66$ corresponds to $2 < \alpha \leq 3$, and thus the mean dispersal distance is not defined (9). An infinite mean in the basic interactions has been observed in diverse systems, and usually implies mean field behavior (18-20). This indicates that each individual virtually interacts with an average environment. Systems showing mean field behavior present a high degree of universality, that is, it is expected that most of the global properties will not depend on field details, such as habitat heterogeneity. This seems to be the case for the English elm spread, since wind dispersion ensures seeds reaching the whole area available as an extreme case of LDD. The fractal dimension of main tumor border calculated from the human glioma invasion was $D = 1.30$ ($R^2 = 0.99$, n = 16). This value corresponds again to predicted power law α exponents for LDD, although to $3 < α ≤ 4$ and thus the mean dispersal distance is defined (9). This is in agreement with the mechanism of glioma cells dispersion, requiring the action of MMPs disintegrating the surrounding extracellular matrix to allow for cell migration and further dispersion along the tumor vessels. This process would set some constraints to the unrestricted LDD allowed by the power law function. Thus, the fractal D value of the patch borders generated by LDD provides information about the degree of interaction of the spreading organism with the surrounding environment.

To further characterize the spatial pattern we calculated the patch size distribution $P(s)$ excluding the initial patch from the simulations with LDD. Patch size is calculated as the number of sites in each patch. $P(s)$ showed a disrupted distribution characterized by two different power laws $P(s) \sim s^{-β}$ at small and large patch areas $s$ separated by a crossover region (Fig. 2A). $β$ value in the small patch areas was 3.37 ($R^2 = 0.98$), while in the large patch areas was 2.24 ($R^2 > 0.98$). Simulations using different values of α produced $P(s)$ curves with similar characteristics (not shown). Tracking individuals in the simulations we determined that the small area section of $P(s)$ is generated by random dispersal and aggregation of reproductively immature individuals. The large area section corresponds to larger patches generated by reproduction of previously dispersed individuals by LDD forming their own patches by localized dispersal and recruitment, originating a founder effect (21) of genetically distinct patches, followed by growth and coalescence of neighboring patches (Movie S1). Large area patches arise only at times greater than first reproduction (not shown). The average minimum area of patches generated by a single individual after its first reproductive event, corresponding to localized dispersion and recruitment, is $s^* = 65 \pm 5$ sites ($\ln(s^*) \approx 4.2$)). These patches, that can be seen in the snapshot from Fig.1A and Movie S1 as small patches with only one reproductive individual, are located at the beginning of the power law corresponding to large patch areas in the $P(s)$ curve (Fig. 2a). $P(s)$ obtained for the glioma invasion (Fig. 2b) was very similar to the simulated patterns. The small and large area power laws show $\beta = 1{,}87$ ($R^2 = 0.97$, n = 4), and $\beta = 1{,}58$ ($R^2 = 0.95$, n = 6), respectively. Remarkably, again s* is located around ln 4, and its value (60 μm$^2$, ($\ln(s^*) = 4.1$)), is compatible with a cluster of approximately 8 cells, indicating the early initiation of a microtumor by localized reproduction from a previously migrated cell. In a similar process, migrated cells of C6 glioma-astrocytoma originated multiple cell groups by division that appeared to be progenitors of tumor masses on Matrigel experiments (22). The English elm field $P(s)$ curve (Fig. 2C) closely resembled the $P(s)$ curves from simulations and glioma invasion. Two power laws appear characterizing the two sections of the curve, with $\beta = 3.27$ ($R^2 = 0.99$, n = 109) for the patches with small area and $\beta = 1.48$ ($R^2 = 0.88$, n = 11) for patches with large areas. We determined s* for English elm as the $s$ value corresponding to the point at the beginning of the second part of the $P(s)$ curve, and found $s^* = 90$ m$^2$ ($\ln(s^*) = 4.5$) (Fig. 2C). This s* value is 3.6 times higher than the minimum field estimated canopy cover of an individual tree at first reproduction and thus compatible with a young patch originated by reproduction from a patch founder parent, followed by localized dispersal given by the SDD component. The scaling pattern of $P(s)$ is remarkable similar to the simulated pattern.

We showed that the LDD mechanism originates similar spatial patterns of spread of cancer cells and a species. This allows for a deeper understanding of dispersal mechanisms in apparently diverse and previously unrelated systems in nature. Specifically, we now know in detail how the LDD process of spread can generate a fractal pattern by patch growth and coalescence, and how a disrupted patch size distribution appears by combination of LDD and localized dispersal and recruitment. We suggest that this is the process driving the early metastatic spread from primary tumors, explained by a particular ecological mechanism of spread. Evidence coming from other fields is supporting this idea. For example, the process of patchy pattern generation we described supports the conjecture of LDD paleocolonization of oak populations occurred 10,000 years ago (2). The existence of patches which are virtually fixed for a single haplotype of chloroplast DNA scattered over several hundred square kilometres (2) can be explained by the LDD process of pattern generation through the founder effect we found. This founder effect is central to the clonal nature of cancer (7), and addresses important questions about clones expansion and control (5). New cell lines disperse far away at very early times from the primary tumor and give early origin to clonal metastatic tumors with resistant characteristics through the founder effect we described. This is particularly relevant to the arising of resistant cell lines during metastatic invasion that challenges the traditional therapeutical approaches (23).

The LDD signature we found is robust since it depends only on the internal population dynamics and dispersal, reflecting self-organizing processes. It appears that it is not fundamentally affected by other ecological processes like competitive interactions and habitat heterogeneity, since we found similar results from the single species, homogenous habitat modeling approach and from the complex field scenario involving the competitive spread of a species into a mountainous forest community (14) and from spread of cancer cells involving interaction with tumor microenvironment (17). A variation appears in the fractal dimension of patch borders in relation to the dispersal environment, from higher D values for the unrestrictive environment (simulation field) to medium and lower values to progressively restrictive environments (native forest and extracellular matrix). In addition, LDD pattern signature is robust to the spatial resolution level of analysis: while the model resolution is maximal (all individuals including newly born ones were traced in the simulations), resolution of real data is lower (only individuals of a minimum detectable size were recorded from the aerial images and cancer cell recognition depended on threshold detection in stained samples).

The choice of the adequate dispersal distribution function, the power law allowing extremely LDD and local subsequent dispersal and recruitment, was crucial both to obtain a sensible and realistic model output and to explain the dynamics of LDD spread. Utilization of bounded or partially bounded distribution functions in modeling LDD process has led to difficulties both in the predictive and explanatory aspects of the models. For example, the use of exponentially bounded dispersal functions for modeling the spread of an invasive moth rendered discrepancies between the observed invasion rates, spatial pattern configuration and fractal characterization (4, 24). In another example, the introduction of a hypothesized chemotactic attraction in a model of brain was necessary to produce a limited cell dispersion around the primary tumor in the simulations (25). In particular, patterns generated by these models did not reflect the spread of the invasive individuals through a large area at very early invasion times. In the case of cancer, early detection is pivotal to improve its treatment or even cure it. Cancers detected at advanced stages are far more likely to cause death than those detected when while the cancer is still confined to the organ of origin (26). Understanding and interpreting the process of cancer invasion in terms of long-range dispersal ecological mechanisms can help to understand the mechanisms of cancer spread and to develop more effective therapeutical strategies.

**Methods**

**The model, numerical simulations and spatial pattern analysis**

We implemented the model as previously described (9, 14, 15). Each site contains in the simulation field at most one individual. At every time step, empty sites are occupied with a probability that depends on the distribution of sites already occupied by mature individuals; the contribution of those sites to the colonization probability is determined by the power law dispersal function allowing for LDD,

$$f(r) = \begin{cases} \dfrac{A}{r^{\alpha}} & \text{if } r \geq \tfrac{1}{2} \\ 0 & \text{if } 0 < r < \tfrac{1}{2} \end{cases}$$

where $A$ is a normalization constant, $r$ is the distance to the parental individual as $r = \sqrt{x^2 + y^2}$ and $\alpha > 2$, as previously described (9). According to the values of $\alpha$ of the LDD power law, when $3 < \alpha \leq 4$ the first moment (the mean) remains finite but the second moment (the variance) becomes infinite. The mean dispersal distance is given by $d \equiv \langle r \rangle = (\alpha - 2) / 2(\alpha - 3)$. When $2 < \alpha \leq 3$ both first and second moments are infinite, and thus the mean dispersal distance is not defined. Finally, for $\alpha > 4$ both the first and the second moments of the distribution are finite. Qualitative similar behaviour can be expected in the global spatial pattern of spread between this last case and SDD (18-20). Simulations began with a single mature individual located at the centre of a square area. Spatial pattern analysis of species spread is based on the statistics of patches of occupied sites and their borders. At a fixed time occupied sites are assigned to patches by giving them a label, representing their corresponding patch number. When an unlabelled occupied site is found, the algorithm creates a new patch by assigning a new label to the current site and to all the connected set of occupied sites associated to it. For each site currently in the cluster all the occupied sites in the set of 8 closest sites are assigned to the same patch. The algorithm continues recursively until no more sites are added to the current patch. The procedure is repeated until no unlabelled occupied sites are left. A patch is then defined as a label that contains more than one site. The border set of a given patch is defined as the list of all the occupied sites lying at its border. For calculating $D$, we plotted the number of boxes $N(l)$ of linear size $l$ as a function of $l$; the fractal dimension is defined as $N(l) \propto l^D$. We calculated the mean fractal dimension $D$ of the patch borders as a function of time, where the averages were taken at fixed times over several simulation runs. We calculated the relative frequency histogram $P(s)$ of patches with size $s$ (excluding the main patch), in an area of 1024 sites at the time when less than 50% of the sites were occupied. After this time the initial patch filled most of the simulation area and few isolated patches remained near the borders of the simulation area. Patch size is defined by the number of sites in each patch.

**Cancer invasion data**

Ang2-promoted human glioma invasion in the brain of animals were determined as described previously (17). Briefly, U87MG Ang2-expressing cell clones ($5 \times 10^5$) were stereotactically implanted into individual nude mouse brains with 5 mice per group. When mice developed neurological symptoms due to disturbance of their central nervous system, mice were sacrificed and their brains were removed, processed and analyzed. The distance of invading glioma cells from tumor masses were assessed by capturing serial images of hematoxylin/eosin-stained brain sections using a Olympus BX51 (Melville, NY) microscope equipped with a SPOT digital camera (Diagnostic Instruments, Inc., Sterling height, MI) and calculated by the fact that under a 100 X magnification, one frame is equal to 1 mm long. Photographs were digitized and interpreted using image processor software, identifying cells at an individual level. Number and area covered for invasive cell clusters generated from the primary tumour were calculated and the fractal dimension of patch borders was calculated using the box-counting method.

**Field species data**

We studied the spatial pattern of spread of English elm using aerial photographs from a forest area of 7 hectares located in a low mountain region of central Argentina. Native forest has been invaded by non-native, competing trees like English elm, Glossy privet (*Ligustrum lucidum*) and Honeylocust (*Gleditsia triacanthos*) (14). English elm is a European tree introduced as ornamental species in the region around the mid 20th century. Reproduction is by seeds. We found no evidence of vegetative reproduction in the field. Individuals bear hermaphrodite flowers. Fruits are samaras released in high numbers and dispersed by wind. Black and white photographs were taken in 1970 (1: 5000), 1987 (1: 20 000), and 1996 (1: 5000). Photographs were digitized and interpreted using image processor software, identifying trees at an individual level. Photographs edges were not used in the interpretation to avoid image distortion. An estimation of error in the photograph interpretation was made by identifying individual trees in the photograph and then checking if they were correctly assigned to the species. In 97 % of cases assignments were correct. In 1970 there were only two near patches of few trees planted, considered as the first dispersal focus in the studied area. Number and area covered of patches generated from the first focus were calculated for 1987 and 1996, and the fractal dimension of patch borders was calculated using the box-counting method.

**Acknowledgments**

This research was supported by grants from Secyt-Universidad Nacional de Córdoba, Agencia Córdoba Ciencia and CONICET (Argentina), the Medical Research Council of the United Kingdom, the National Institute of Health, USA (CA102011), American Cancer Society, USA (RSG CSM-107144) and the Hillman Fellows Program (S.-Y. C. and B. H.). D.E.M. and S.A.C. are members of the National Research Council (CONICET, Argentina). We acknowledge Dante Chialvo for a critical reading of the original manuscript.

**Figure legends**

**Fig. 1.** Spatial patterns of spread with long-distance dispersal from simulations and real data are similar. (a) Simulation spread from an individual in an area with L = 320 sites after 9 years from first reproductive time. Reproductive (black dots) and immature (grey dots) individuals are shown. Power law LDD with $\alpha = 3.11$. (b) Spatial spread of human invasive glioma in (black areas) in mouse brain (white ground). (c) Spatial spread of English elm (black areas) surrounded by a mixed forest of other invasive and native species (white ground).

**Fig. 2.** The size distribution of patches $P(s)$ from simulations and real data curves are characterized by power laws ($P(s) \sim s^{-\beta}$), corresponding to patches of small and large areas. (a) $P(s)$ for long-distance simulations for $\alpha = 3.33$…$s$ is given in number of sites. (b) $P(s)$ for spread of human invasive glioma. s is given in $\mu m^2$. (c) $P(s)$ for field spread of English elm. $s$ is given in $m^2$. $\beta$ values are given in the text. $s^*$ indicates the estimated average minimum size of a patch generated by a single individual by reproduction and localized dispersal in the simulations (a) and in real data (b,c).

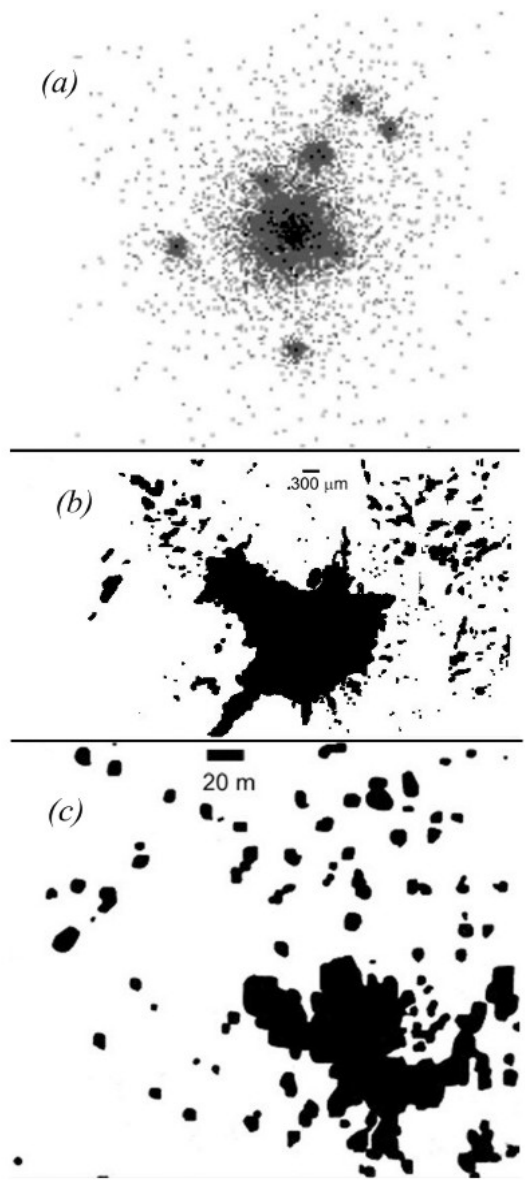

**Fig. 1**

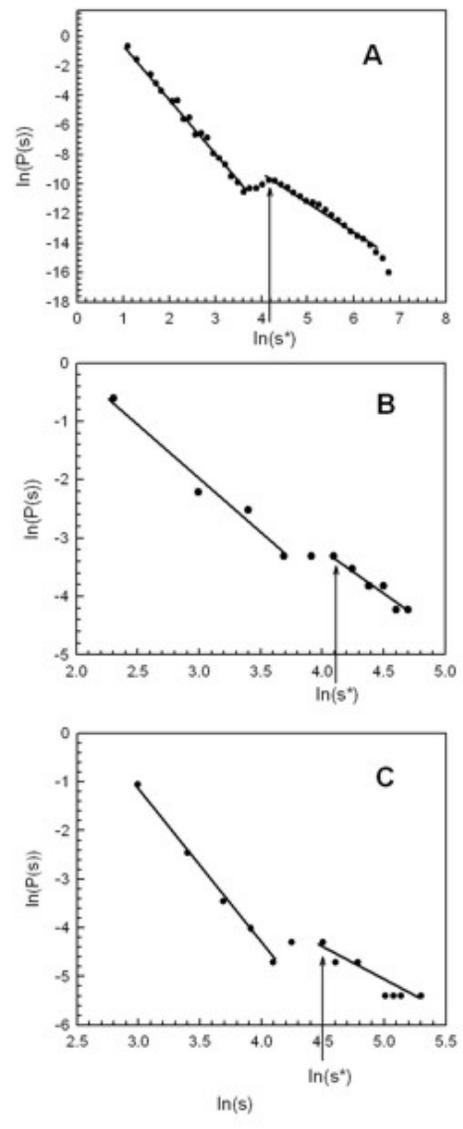

**Fig. 2**